\documentclass[12pt,a4paper]{article}

\usepackage[english]{babel}

\usepackage{epsfig}
\usepackage{dcolumn}
\usepackage{bbm}      
\usepackage{amsmath}
\usepackage{cite}
\usepackage{changebar}
\usepackage{graphicx}

\makeatother 

\renewcommand{\thefootnote}{\arabic{footnote}}

\newcommand{\Tr} {\mbox{Tr}}

\begin{document}

\thispagestyle{empty}
\hbox{}
 \mbox{} \hfill \hspace{1.0cm}

 \mbox{} \hfill BI-TP 2003/18\hfill

\begin{center}
\vspace*{1.0cm}
\renewcommand{\thefootnote}{\fnsymbol{footnote}}
{\huge \bf ENTROPY FOR COLORED} \\ ~\\
{\huge \bf QUARK STATES AT}\\ ~\\
{\huge \bf FINITE TEMPERATURE~}\\

\vspace*{1.cm}
{\large David E. Miller$^{1,2}$} and 
{\large Abdel-Nasser~M.~Tawfik$^{1}$}\\
\vspace*{.5cm}
${}^1$ {Fakult\"at f\"ur Physik, Universit\"at Bielefeld, Postfach 100131,\\ 
D-33501 Bielefeld, Germany
\footnote{email: dmiller@physik.uni-bielefeld.de; tawfik@physik.uni-bielefeld.de}}\\
${}^2$ {Department of Physics, Pennsylvania State University,
Hazleton Campus,\\
Hazleton, Pennsylvania 18201, USA 
\footnote{email: om0@psu.edu} }\\
\vspace*{.5cm}

{\large \bf Abstract \\ }
\end{center}
The quantum entropy at finite temperatures is analyzed by using models for
colored quarks making up the physical states of the hadrons. We explicitly
work out some special models for the structure of the states of $SU(2)$ and
$SU(3)$ relating to the effects of the 
temperature on the quantum entropy. We show that the entropy of the
singlet states monotonically  decreases meaning that the mixing of these
states continually diminishes with the temperature. It has been found that
the structure of the 
octet states is more complex so that it can be best characterized by two
parts. One part is very similar to that of the singlet states. The other
one reflects the existence of strong correlations between two of the three
color states. Furthermore, we work out the entropy 
for the {\it classical} Ising and the {\it quantum} $XY$ 
spin chains. In Ising model the quantum (ground state) entropy does not directly 
enter into the canonical partition function. It also does not depend on the
number of spatial dimensions, but only on the number of quantum states
making up the ground state. Whereas, the $XY$ spin chain has a finite
entropy at vanishing temperature. The results from the spin models
qualitatively analogous to our models for the states of $SU(2)$ and
$SU(3)$.

\vfill
\noindent 
PACS:         05.30.-d \hspace*{3mm} Quantum Statistical Mechanics, \\
\hspace*{13mm}12.39.-x \hspace*{3mm} Phenomenological Quark Models, \\
\hspace*{13mm}03.75.Ss \hspace*{3mm} Degenerate Fermi Gases,
\newpage

\section{\label{sec:1}Introduction}

In this paper we present a new method for evaluating the entropy of the
colored quark states at finite temperature. Clearly, the entropy which we
mean here is the quantum entropy which is to a great extent different in
nature from the {\it classical} (Boltzmann-Gibbs) entropy. From the
classical point of view the entropy characterizes an access to information
between macroscopic and  microscopic physical quantities using statistical
criteria. It can be considered to be the number of possible microstates 
that the macrosystem can include. Whereas, the quantum
entropy\footnote[1]{The authors of \cite{europhys1} 
distinguished between different entropies: the classical one they called the
measurement entropy, and their relevant entropy refers to our quantum
entropy.} 
can be calculated for very few degrees of freedom by using the density
matrix and should not vanish for zero temperature. It reflects the uncertainties 
in the abundance of information about the quantum states in a system. The
definition of quantum entropy may date back to works of von Neumann~\cite{vNeu}
in the early thirties on the mathematical foundations of quantum theory. Another
milestone has been set out by Stratonovich,
when he introduced the reciprocative quantum entropy for two
coupled systems. In the Lie algebra the quantum fluctuation and correspondingly
the quantum entropy are given by non-zero commutative relations. 
The most essential information we can retain from the physical systems is
similar to measuring the eigenvalues of certain observables. The faultless
measurement is to be actualized only if the degeneracy in the determined
observable is entirely considered.\\   
~\\
\indent
     According to the Nernst's heat theorem the entropy of a homogeneous
     system at zero temperature is expected to be zero. However, it has been
proven that the mixing in ground states of {\it subsystems} with conserving
the particle number, the temperature and the volume results in   
a finite entropy~\cite{Planck64}, the so-called Gibbs paradox. On the other
hand, we know that the 
effects of quantization on the fundamental laws of thermodynamics were
generally well known to the founders of quantum theory~\cite{LaLi}. Planck, who
had successfully predicted the measured intensity distribution of different
wavelengths by the postulation of a minimum energy size for the light
emitted from a dark cavity at a given temperature, 
had also realized that the massive particles
with non-zero spin also possess a finite entropy relating directly
to that of the spin in the limit of low temperatures. This 
realization provided the chance for his students to resolve 
the Gibbs paradox, 
furthermore, it also offered an exception to Nernst's heat 
theorem, which is usually stated under the name of
the third law of thermodynamics that the entropy
of a closed system in equilibrium must vanish in the
low temperature limit. A clear general discussion of the 
entropy relating to Nernst's heat theorem was given by
Schr\"odinger some years later in a lecture series \cite{Schr}. In his
selected example Schr\"odinger took an $N$-particle-system, in which
each one has two quantum states contributing to the  
many particle ground state. Thus the ground state held $2^N$
degenerate configurations in its structure, which must then 
provide for it an entropy of $N\ln2$. This value is 
obviously independent of all the thermodynamical quantities 
other than the number of particles $N$ itself. 

     Furthermore, it is known that the more information present, the greater
is the reduction of the entropy. In other words the entropy is reduced
by the amount of information distinguishing the different degenerate states of the 
system. The completely mixed state is that of minimal information. This is
completely consistent with the idea of confinement in the hadronic matter. If
we were to consider a gas of $N$ hadrons in the sense that
Schr\"odinger~\cite{Schr} considered a gas of two level atoms, 
we should roughly expect an entropy of the form $N{\ln 3}$ for the
$SU(3)_c$ quark structure in the color singlet ground state of this pure
composite system~\cite{Mill}. In this article we attempt to elucidate the
physical meaning of the quantum entropy for the colored quark states and
investigate its behavior at finite temperature. 

     According to the third law of thermodynamics it is expected that the
confined hadron bags - as a pure state - have zero entropy at zero
temperature. But the hadron constituents are to be treated as  
subsystems composed of quantum elementary particles, which in their 
ground states can exhibit a finite value of the entropy. 
The quantum entropy of such subsystems reflects the degree of mixing 
and entanglement~\cite{vidal} inside the hadron bag. 
In a recent letter~\cite{Mill} it was shown that in a quantum
system with $SU(3)_c$ internal color symmetry how the ground state
entropy arises in relation to the mixing of the quantum color states 
of that system. So far we conclude that in a completely mixed ground state
with $N$ internal components the entropy relates directly to the value $\ln N$.  
Nevertheless, one would expect some changes in this entropy
in the presence of other states or at finite temperatures.
Such changes were mentioned in \cite{Mill} in relation to the octet hadronic
states and the changes in the ground state entropy at finite
temperature.\\ 
~\\
\indent
     Understanding the thermal behavior of quantum subsystems \cite{Feyn1} -
in our case the colored quark states - is very useful for 
different applications. It may well bring about a device 
for the further understanding of the recent lattice results with 
the heavy quark potential~\cite{Felix1,Felix2,FelixPhD}. The lattice results for the
entropy difference in a quarks-antiquark singlet state lead to the value of
$2\ln 3$ at vanishing temperature~\cite{Felix2,FelixPhD}. 
Our evaluation for the quantum entropy of one quark as part of the
colorless singlet ground state~\cite{Mill} yields the value of $\ln3$. 
Furthermore, we believe that the investigation of the quantum subsystems at
finite temperature might be useful for 
understanding the concept of confinement, which could exists everywhere
throughout \hbox{$T\in[0,\infty]$}. The cold dense quark matter in the
interior of stellar compact objects provides another application for this
work. Also we think that according to the existence of finite entropy
inside the hadron bags, the confined quarks get an additional heating and the
bag constant should be correspondingly modified. The quark distributions inside the
hadron bags reflect themselves as entanglement or - in 
our language - quantum entropy. \\   
~\\
\indent
     In this work we shall further investigate the  
implications of the quantum entropy for the colored quarks relating 
to the known properties of the standard model~\cite{DoGoHo}.
As previously mentioned~\cite{Mill} we shall not bring in
other important properties of the quarks in the standard
model like flavor, spin, isospin, chirality, electric charge
and the spatial distribution of the quarks. From now on in
this work we shall assume that we are only looking at the 
color symmetry so that we shall leave out any name 
distinctions between the states. Here we shall further look
into some specific properties of the other quark structures.
Thereafter, we shall investigate some specific properties 
of the other states in relation to the ground state, 
which lead to some more results for the entropy. As a following 
investigation we propose and solve for the quark quantum entropy 
some simple models of color unmixing at finite temperatures, 
for which we check each model for the limiting cases.
Then we study the correlations of entangled spin systems. These spin
systems are used to compare our results for measurement entropy of a system
of colored quark states. Other degrees of freedom are not considered. \\

     This paper is organized as follows: the next section is devoted to the
formulation of the ground state, from which we develop models for the entropy of
mixing of colored quark $SU(2)$ and $SU(3)$ states. Then we introduce our
thermal models for the quantum entropy of these states. Some spin models
are investigated in  relation to the known exact solutions. The following
section contains the discussion of our results. Finally, we end with the
conclusion and outlook.

\section{\label{sec:2}Formulation for the Ground State Entropy}

In order to understand the ground state structure, we recall some common
properties of spin and color quantum systems which relate to the ideas of
superposition and entanglement~\cite{NieChu}. The orthonormal basis usually
taken~\cite{Feyn1} for spin  $SU(2)$ can be written as $|0\rangle$ and
$|1\rangle$ for the two states. For such two-state-system we have four
combinations of $|ij\rangle$
a useful linear orthonormal
combination thereof. When we use the Pauli matrices $\sigma^x,
\sigma^y, \sigma^z$ together with the two dimensional identity
matrix ${\mathbbm{1}}_2$, we may easily write down the singlet
and triplet structure for the structure of the states of $SU(2)$.
The usual symmetric triplet and antisymmetric singlet states
also provide a proper basis. After we have written down the density
matrices $\rho_{t}$ and $\rho_{s}$ for each of the states, we find 
that after projecting out the second state the single quark reduced 
density matrices are both in the form

\begin{equation}
\label{eq:denmat2}
          {\mathbf{\rho}}_{q,2} = \sum_i p_i {\mathbbm{1}}_2 
\end{equation}

\noindent
where $p_i$  is the probability of $i$-th state. We can calculate the
entropy $S$ of the quantum states~\cite{vNeu,LaLi}, which makes direct 
use of the density matrix $\bf{\rho}$ 

\begin{equation}
\label{eq:entropy}
                   S = -\Tr\; (\bf{\rho}\; \ln{\bf{\rho}}),
\end{equation}

\noindent
Nevertheless, this equation could be found in text books, we apply it
here where the trace is taken over all the quantum states
(Eq.~\ref{eq:denmat2}). For quantum  
operators the trace is independent on the representation, therefore the quantum
states might well be used to write down the either quantum canonical or grand
canonical partition function.  The density matrix can be illustrated as 
mixing of {\it sub}systems of a closed system. According to the Nernst's heat
theorem this enclosed system has zero entropy at zero
temperature~\cite{Planck64}.

\noindent
When, as is 
presently the case, the eigenvectors are known for $\rho$, we may directly 
write this form of the entropy in terms of the eigenvalues ${\lambda}_i$
as follows: 

\begin{equation}
\label{eq:enteig}
                   S = - \sum_i \lambda_i \; \ln \lambda_i
\end{equation}

\noindent
It is obviously important to have positive eigenvalues. For the special 
case of a zero eigenvalue we use the fact that $x{\ln x}$ vanishes in the 
small $x$ limit. Then for the density matrix $\rho$ we may interpret 
${\lambda}_i$ as the probability  $p_i$ of $i$-th state. 
This meaning demands that $0<p_i\leq 1$. Thus the orthonormality 
condition for the given states results in the condition
This is a very important condition for the entropy.
Thus we can easily see that for $SU(2)$ the value of probabilities 
$p_i$ is always $1/2$ yielding the same total entropy
for both the singlet and triplet states

\begin{equation}
\label{eq:entetwo}
                   S_{q,2} = \ln 2
\end{equation}

We get the same results if we apply the so-called Schmidt decomposition
on the pure state $|\psi>$, which consists of {\it sub}states, $|0>$
and $|1>$, then there is sets of orthonormal states, $\{|0>_i\}$ and
$\{|1>_i\}$, so that  
\begin{eqnarray}
|\Psi> &=& \sum_i {\cal N}_i\;|0>_i \otimes |1>_i \label{schmidt1}
\end{eqnarray}
where ${\cal N}$ are real position numbers, known as the Schmidt numbers,
which satisfy
\begin{eqnarray}
\sum_i {\cal N}_i^2 &=& 1 \label{schmidt2}
\end{eqnarray}
One important consequence of last equation, is that the reduced density
matrices of {\it sub}states, $|0>$ and $|1>$ should have identical
eigenvalues. Plugging ${\cal N}_i\equiv\lambda_i$ in Eq.~\ref{eq:enteig},
$S_{q,2}=\ln2$. Eq.~\ref{schmidt1} can be generalized for decomposition
into $n$ subsystems. The pure state $|\psi>$ can be expanded in a number of 
factorizable $n$-states. The number of coefficients is minimal and
Eq.~\ref{schmidt2} still valid.  \\

     In the case of $SU(3)_c$ the state structures for the singlet 
and octet are very different. We recall that we have found
for the single quark density matrix $\rho_q$ the form~\cite{Mill}

\begin{equation}
\label{eq:denmat3}
          {\mathbf{\rho}}_{q,3} = \sum_i p_i \; {\mathbbm{1}}_3
\end{equation}

\noindent
where ${\mathbbm{1}}_3$ is the three dimensional identity matrix. \\

\noindent
We now apply the above definitions of the entropy to the $SU(3)_c$
quark states -- as was done in~\cite{Mill}.
It is clear that the original hadron states are pure colorless states
which posses zero entropy as in the third law of thermodynamics. For the 
meson it is immediately obvious since 
each colored quark state has the opposing colored antiquark state for
the resulting colorless singlet state. The sum of all the cycles determine
the colorlessness of the baryon singlet state thereby giving no entropy.
However, the reduced density matrix for the individual quarks (antiquarks)
${\rho}_q$ or ${\rho}_{\bar{q}}$ has a finite entropy. In this context, the
reduced density matrix can be illustrated as a certain mixing inside the closed
system~\cite{Planck64}. Meanwhile the whole system have zero entropy at
zero temperature, the {\it sub}systems - as shown above - are expected to
have finite entropy at vanishing temperature.  \\
~\\ 
\noindent
     For $SU(3)_c$ all the eigenvalue ${\lambda}_i$ in Eq.~\ref{eq:enteig} have
the same value $1/3$. Thus we find for all quark (antiquark) in  singlet states

\begin{equation}
\label{eq:entquark}
                  S_{q,3} = \ln 3
\end{equation}

\noindent
As a further exercise we may compare this result with those
of the quark octet states. The octet density matrices ${\rho}_{o,i}$ may
be constructed from the eight Gell-Mann matrices $(\lambda)_i$ with 
$i=1,2,\cdots,8$. The density matrix for each state is constructed by
using the properties of ${\Psi}(\lambda)_i{\Psi}^{*}$.
The first seven matrices all give the same value for the 
entropy $\ln 2$, since all of these states are constructed only from
the Pauli matrices. This result comes from the fact that these first seven
octet states each involve only two of the three color states -- that is
these octet states are not pure states in all the colors. Although the 
mixing of the two states is equal, it is not complete since the third 
color is absent. However, the eighth diagonal Gell-Mann matrix involves
all three colors, but the mixing is unequal. It yields an entropy

\begin{equation}
 ~S_{o,8} = \ln 3 - {\frac{1}{3}}\ln{2}
\label{eq:enteight}
\end{equation}

\noindent
Thus we can clearly state that the entropy of any of the quark  octet states 
is always smaller than the quark color singlet state. 
This means that the colorless quark singlet state is the most probable 
individual state for the hadrons.

\section{\bf Thermodynamical Models for Mixed Colored Quark States}

The structure of the color singlet hadronic ground state for $SU(3)$ was
shown in~\cite{Mill} to have a complete uniform mixing of all  
the colors of the quarks and antiquarks for both the mesons and the baryons. 
We have seen above that this situation does not happen for the single 
quark (antiquark) entropy $S_q$ or $S_{\bar{q}}$ in the presence of octet 
states, where the mixture is either partial or unequal.
We now want to return to the single quark (antiquark) reduced density
matrices ${\mathbf{\rho}}_{q,2}$ or ${\mathbf{\rho}}_{\bar{q},2}$,
which we shall assume to be the same for the fundamental and antifundamental
representations since we are not considering the differences between flavors.
In this section we will extend these calculations of the structural entropy
of the ground states for the $SU(2)$ and $SU(3)$ quarks in color singlet 
states to models for color mixing at finite temperature. In these models
the Boltzmann weighting for the finite temperature states of the single
colored quark states will contain the single particle relativistic
energies $\varepsilon(p)$, which is given by $\sqrt{m^2~+~p^2}$ for the
relativistic quark momentum $p$ and mass $m$. Since the biggest effect at the given
temperature $T$ comes with the lowest value of $\varepsilon(p)$, 
we may well assume that $\varepsilon(p)$ is just determined by the 
lowest quark mass threshold. 
Furthermore, $\varepsilon(p)$ is also color independent.

\subsection{\label{subssu2}$SU(2)$ Thermodynamical Model}

\begin{figure}
\vskip -.2cm
\begin{minipage}[t]{49mm}
\includegraphics[width=8.cm,]{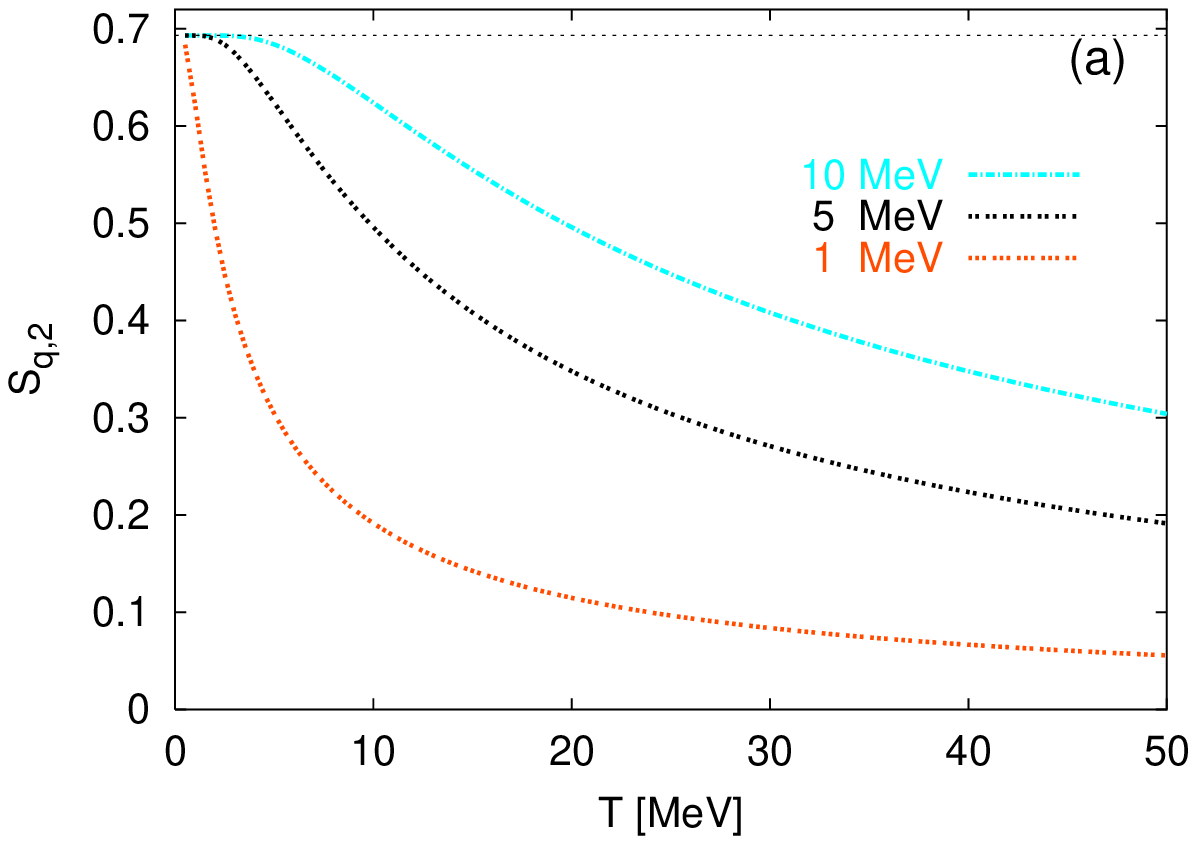}
\end{minipage}
\hskip 0.4cm
\begin{minipage}[t]{7.cm}
\hspace*{2.2cm}
\includegraphics[width=8.cm]{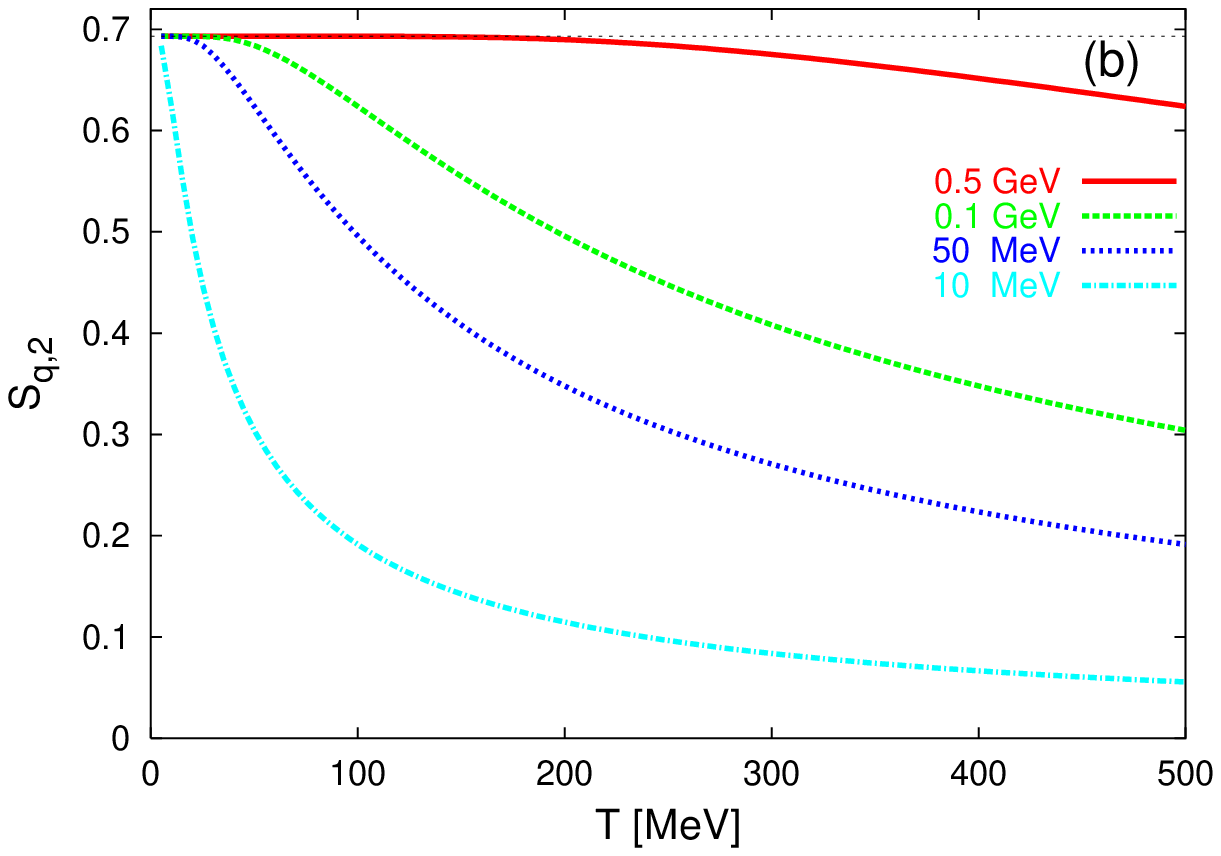}\\
\end{minipage}
\vskip -0.5cm \caption{\label{fig:1}\sf \footnotesize
  Single quark entropy for various
  quark masses as a function of the temperature $T$ for $SU(2)$,
  Eq.~\ref{eq:entrotemp2}. The left panel shows the 
  results for the stated light quark masses. The right panel
  depicts the results for masses up to $500\;$MeV. 
  The dotted lines at the top represent the value of the
  ground state entropy at zero temperature in Eq.~\ref{eq:entetwo}. 
  }
  \end{figure}

     We start our consideration of thermodynamics with a very simple model
for quarks with an internal $SU(2)$ symmetry at a finite temperature $T$.
We postulate that the single quark reduced density matrix with two colors
have the following form:
\begin{eqnarray}
\label{eq:denmattem}
{\mathbf{\rho}}_{q,2}(T) &=& \frac{1}{2}\left[\left(1-e^{-
                         \varepsilon(p)/T}\right)|0><0| + \left(1+e^{-
                         \varepsilon(p)/T}\right)|1><1|\right] \nonumber \\
                         &=&\frac{1}{2} \left[{\mathbbm{1}}_2 
                           -\sigma^z {e^{- \varepsilon(p)/T}}\right]
\end{eqnarray}
\noindent
which shifts the weighting of the eigenstates due to the temperature $T$. 
Thus we note that the total probability is still one, so that
${\mathbf{\rho}}_{q,2}(T)$ will still have the correct probabilistic
interpretation. Furthermore, we note that if we had taken all the Pauli
matrices with equal weighting in front of the Boltzmann factor, we would
still be able to diagonalize the full $SU(2)$ reduced density matrix into
this form. Then the eigenvalues from the $T$-depending reduced density matrix,
Eq.~\ref{eq:denmattem}, read 
\begin{eqnarray}
\det\left[<i|\rho_q|j>-\frac{\lambda}{2}\right] &=&
   \left[\frac{1}{2}-\frac{e^{-\varepsilon(p)/T}}{2\sqrt{3}}-\lambda\right]
   \left[\frac{1}{2}+\frac{e^{-\varepsilon(p)/T}}{2\sqrt{3}}-\lambda\right]
   -\frac{e^{-2\varepsilon(p)/T}}{6} = 0 \nonumber \\
\lambda_{i}(T) &=& \frac{1}{2}\left(1-\sigma^z e^{-\varepsilon(p)/T}\right)
\end{eqnarray}
$i=\pm1$ refers to the states, $|0>$ and $|1>$, which are included in $\sigma^z$ in
Eq.~\ref{eq:denmattem}. 

\noindent
As it was above carried out for the ground state, we are now
able to calculate the entropy $S_{q,2}(T)$ at finite
temperature~\cite{vNeu,LaLi} 
from the eigenvalues $\lambda_i(T)$ and under the assumption that the color
at finite temperature have the same energy eigenstates, we find
\begin{eqnarray}
S_{q,2}(T) &=& - \sum_{i=\pm1} \lambda_{i}(T) \ln \lambda_{i}(T)
           \label{eq:entrotemp} \\ 
           &=& -\frac{1}{2}\left(1 - e^{-\varepsilon(p)/T} \right)
                \ln\left[\frac{1}{2}\left(1 - {e^{-\varepsilon(p)/T}}\right)\right]
                \nonumber\\
           & & -\frac{1}{2}\left(1 + {e^{-\varepsilon(p)/T}}\right)
     \ln\left[\frac{1}{2}\left(1 + {e^{-\varepsilon(p)/T}}\right)\right]
\label{eq:entrotemp2}
\end{eqnarray}

In the low temperature limit we can immediately find that the ground state
quark entropy $S_{q,2}(0)$ has again, as in Eq.~\ref{eq:entetwo}, 
the value of $\ln2$, a completely equally mixed quark color state. However in the
high temperature limit the contribution to the entropy of the first state
vanishes from deoccupation, while the second state becomes a pure state with
a probability of one, which also contributes a vanishing value to the entropy.
In Fig.~\ref{fig:1} we look at some particular cases for
$S_{q,2}(T)$ both  
at lower and higher temperatures. In Fig.~\ref{fig:1}a we see how the single
quark entropy varies in a range of temperature up to $50\;$MeV with
quark masses $1$, $5$ and $10\;$MeV. We notice that in all cases the entropies
monotonically decline in this region meaning that the mixing of the states
continually decreases. For the sake of
comparison in Fig.~\ref{fig:1}b we look 
at some larger quark masses in a much larger temperature range. We see the 
same tendency for the separation of the states. We note that 
with increasing quark masses the range of temperature within which the
entropy is entirely determined by the ground state value becomes wider 
and the entropy remains larger for higher temperatures. This reflects the
importance of the ground state entropy for the massive quark systems. 
Thus in this model for $SU(2)$ colored quark states we find in the high
temperature limit a pure single colored quark state in the classical
sense. Whereupon, we would expect a pure quark phase in which all the
correlations between  the different colors have vanished - free quarks!

\subsection{\label{subssu3}$SU(3)$ Thermodynamical Model}

\begin{figure}
\vskip -.2cm
\begin{minipage}[t]{49mm}
\includegraphics[width=8.cm]{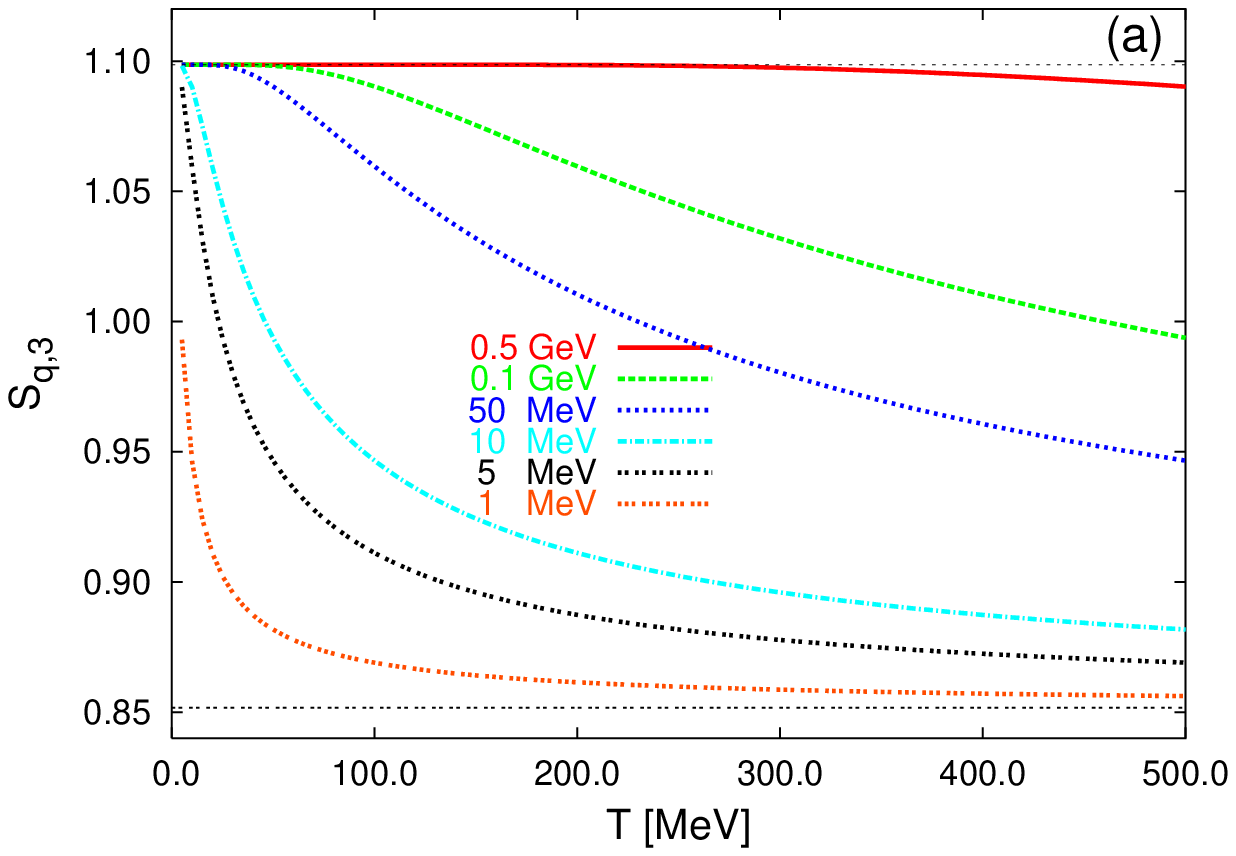}
\end{minipage}
\hskip 0.4cm
\begin{minipage}[t]{7.cm}
\hspace*{2.2cm}
\includegraphics[width=8.cm]{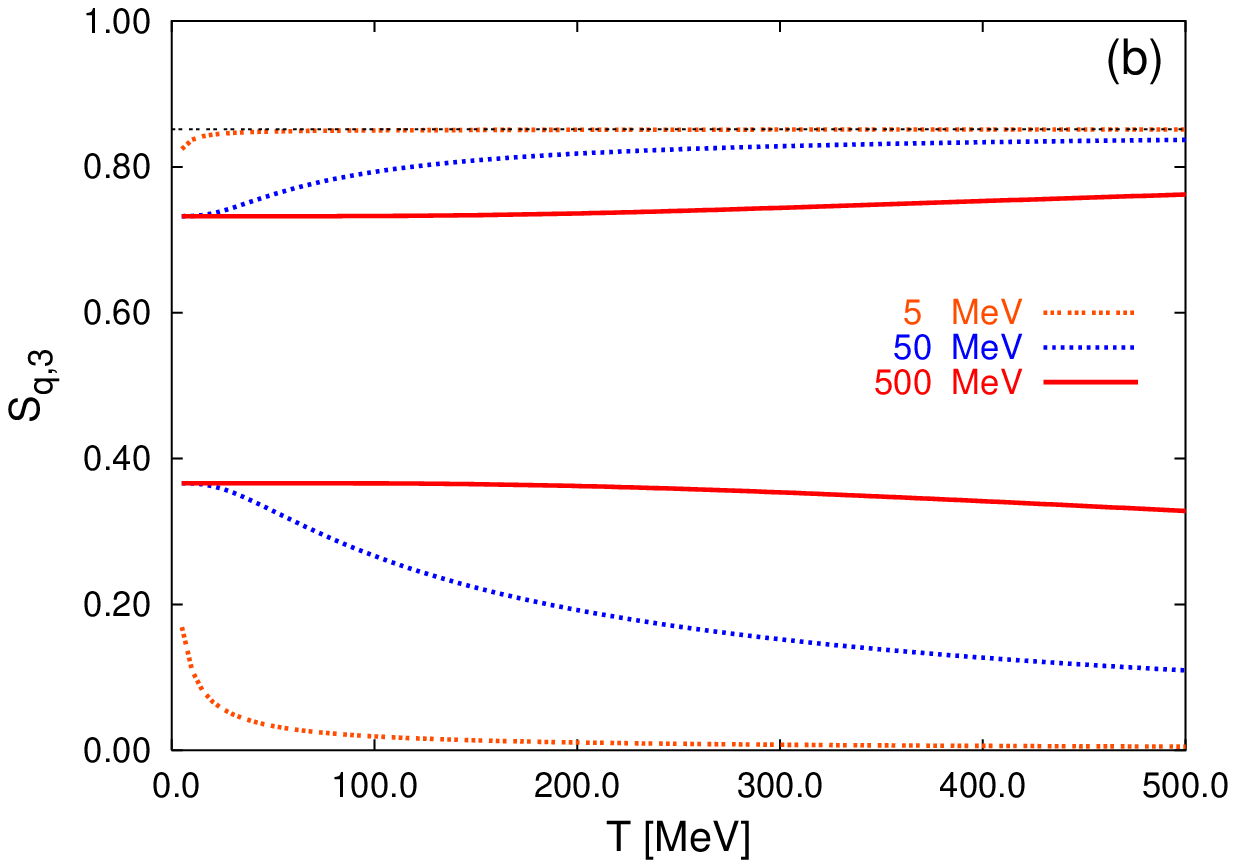}\\
\end{minipage}
\vskip -0.5cm 
\caption[Fig. 2]{\label{fig:2}\sf \footnotesize
  The left panel shows the single quark entropy
  change for various quark masses as a function of temperature $T$ for $SU(3)$,
  Eq.~\ref{eq:sq3}.  
  The right panel gives details on this behavior. We plot separately the two 
  contributing terms of Eq.~\ref{eq:sq3} for masses, $1$, $50$ and 
  $500\;$MeV. The curves in bottom part of the figure represent only the first
  term.  }
\end{figure}

     Now we construct a similar model for $SU(3)$ with the right quark 
mixing in the ground state, which also provides the proper probabilities
for each of the states from the trace condition $\Tr {\rho}={\sum}_i p_i=1$ (section~\ref{sec:2}). 
Again we demand that the energy eigenstates remain the same for each 
of the color states  $|i\rangle$. We think that the eighth (diagonal) Gell-Mann
matrix $\lambda_8$ is not a very suitable choice as a weighting matrix for
the thermal states even though it maintains the trace condition for the 
probability. The eighth Gell-Mann matrix weights the third
color twice as much as the other two taken individually. Therefrom, we
should look for another weighting matrix for the thermal states of 
$SU(3)$. A good possibility, which we shall choose, is that we take the 
three complex roots of $-1$, which are ${\cal R}_0=-1$, ${\cal
  R}_1=\exp{i\pi/3}$ and ${\cal R}_2=\exp{-i\pi/3}$. 
\begin{eqnarray}
\rho_{q,3} &=& \frac{1}{3}\left[{\mathbf 1}_3+\left({\cal R}_0|0><0|+{\cal
      R}_1|1><1|+{\cal R}_2|2><2|\right)e^{-\varepsilon(p)/T}\right]
      \label{rhoq3:1}
\end{eqnarray}
Then we get three eigenvalues:
\begin{eqnarray}
\lambda_0 &=& \frac{1}{3}\left(1-e^{-\varepsilon(p)/T}\right),\nonumber \\ 
\lambda_1 &=& \frac{1}{3}\left(1+e^{i\pi/3}\; e^{-\varepsilon(p)/T}\right),\\
\lambda_2 &=& \frac{1}{3}\left(1+e^{-i\pi/3}\; e^{-\varepsilon(p)/T}\right)\nonumber
\end{eqnarray}
We may write the three roots as the weights $w_i$ for 
the states $|i\rangle$ with the Boltzmann factor. Thus the reduced quark
density matrix, Eq.~\ref{rhoq3:1}  for $SU(3)$ reads 
\begin{equation}
\label{eq:denmattem3}
 {\mathbf{\rho}}_{q,3}(T) = \frac{1}{3} \left({\mathbbm{1}}_3 
           + |i\rangle\; w_i\; {e^{-\varepsilon(p)/T}}\; \langle i| \right).
\end{equation}

\noindent
Clearly, in the low temperature limit we get back the completely mixed 
state with a probability of $1/3$ for each color state $|i\rangle$.
We are able to calculate the entropy $S_{q,3}(T)$, Eq.~\ref{eq:entrotemp}
 by carefully using the proper definitions for the complex logarithms. 
\begin{eqnarray}
S_{q,3}(T) &=& -\frac{1}{3}\left(1-{e^{-\varepsilon(p)/T}}\right)\;
                \ln\left[\frac{1}{3}\left(1-{e^{-\varepsilon(p)/T}}\right)\right]
                \nonumber\\
           & &
                -\frac{1}{3}\left(1+e^{i\pi/3} e^{-\varepsilon(p)/T}\right)\;
                \ln\left[\frac{1}{3}\left(1+e^{i\pi/3}
                e^{-\varepsilon(p)/T}\right)\right]  \nonumber \\
           & &
                -\frac{1}{3}\left(1+e^{-i\pi/3} e^{-\varepsilon(p)/T}\right)\;
                \ln\left[\frac{1}{3}\left(1+e^{-i\pi/3}
                e^{-\varepsilon(p)/T}\right)\right]  
\end{eqnarray}
Using the properties of the logarithms of complex
variable\footnote[2]{ For the phase $ -\pi<\theta\leq+\pi$ we define the complex
  variable as ${\cal Z} = {\cal R}e Z\; e^{i \theta}$.} we can write the
real part as 
\begin{eqnarray}
z &=& {\frac{1}{3}} \left[1 + e^{-\varepsilon(p)/T} + e^{-2 \;
                \varepsilon(p)/T} \right]^{1/2} \label{eq:defz} 
\end{eqnarray}
and the phase is given by
\begin{eqnarray}
\theta &=& \arctan{\left(\frac{\sqrt{3} \; e^{-\varepsilon(p)/T}}
          {2 + e^{-\varepsilon(p)/T}}\right)} \label{eq:deftheta}
\end{eqnarray}
After a little algebra we find that the single quark quantum entropy at
finite temperature for $SU(3)$ reads 
\begin{eqnarray}
S_{q,3}(T) &=& -\frac{1}{3}\left(1 - {e^{-\varepsilon(p)/T}}\right) 
                \ln\left[\frac{1}{3}\left(1 - {e^{-\varepsilon(p)/T}}\right)\right]
                \nonumber\\
           & & - 2 z \left[\ln(z) \cos(\theta) - \theta\; \sin(\theta) \right],
                \label{eq:sq3} 
\end{eqnarray}

\noindent
In the low temperature limit we have $z$ just equal to $1/3$ and
 $\theta$ is exactly zero. Then the quark entropy is clearly again just $\ln 3$.
However, in the limit of very high temperature $z$ becomes $1/\sqrt{3}$
and $\theta$ is just $\pi/6$. The contribution of the first term of
 Eq.~\ref{eq:sq3} to $S_{q,3}(T)$ simply vanishes as was the case for
 $S_{q,2}(T)$ (review Fig.~\ref{fig:1}). However, the second term 
still remains at high temperatures leaving a limiting entropy of $0.8516$.
We see this effect in Fig.~\ref{fig:2} where we have plotted $S_{q,3}(T)$
 for various values of the quark masses.
We note that the first term behaves qualitatively very similarly to
 the single quark for $SU(2)$  
in Fig.~\ref{fig:1}. The entropy begins from $\ln 3/3$ and exponentially
decreases with the temperature $T$. The second term has a remarkable dependence 
on $T$. For $T\rightarrow 0$, it has the value of $2/3\; (\ln 3)$ analogously to
 $SU(2)$. 
With increasing temperature it rapidly goes to the asymptotic value
$(\ln 3)/2 + \pi(\sqrt{3}/18)$. Furthermore, we note that
by decreasing the mass it limits the range of temperature 
to reach this asymptotic region. Since the asymptotic value at high
 temperatures has  remained more than three quarters 
of its ground state value of $\ln 3$, there are still considerable correlations
between two of the three color states. This observation points to an important
fact about the structure of $SU(3)$ in these statistical models:  the root
structure forbids a complete cancellation of the real solutions which is needed
to maintain the trace condition on the reduced density matrix. Thus two states
are always matched against one. Hence for this model in the high temperature
limit one color vanishes while the other two remain mixed and thereby
correlated. The ground state favors the color singlet state with complete
mixing. In this model the high temperature limit favors the octet states involving
mostly two colors. This situation for $SU(3)$ can be contrasted with $SU(2)$
where the triplet and the singlet states have the same reduced density matrices
except for the pure triplet states.

\section{\bf Spin Models with Strong Correlations}

Our objective in this section is to further investigate the
structure of the entropy for some known 
spin models at finite temperature in which strong correlations exist. 
The general category of all spin models goes under
the name of the Heisenberg model, which is the central to the theory
of magnetism~\cite{Yos}. In general, the Hamiltonian for these models involves
the vector of spin matrices ${\bf s}{({\bf r}_i)}$ for an electron
at the position ${\bf r}_i$ coupled to another electron  with  
the vector of spin matrices ${\bf s}{({\bf r}_j)}$ 
at position ${\bf r}_j$

\begin{equation}
{\mathcal{H}} = -2 \sum_{i<j}  J_{ij}\; {\bf s}_i \cdot {\bf s}_j,
\label{eq:Heisham}
\end{equation}

\noindent
where $J_{ij}$ is the exchange integral arising from the integration
over the interaction containing the overlap of the spatial wavefunctions
of the two electrons located at the two points ${\bf r}_i$ and ${\bf r}_j$.
The relation of this exchange integral $J_{ij}$ to the different spin
directions ${\bf s}_i$ and ${\bf s}_j$ determines the local structure
of the interaction.

\subsection{Classical Ising spin chain}

\begin{figure}
\vskip -.2cm
\begin{minipage}[t]{80mm}
\includegraphics[width=8.cm]{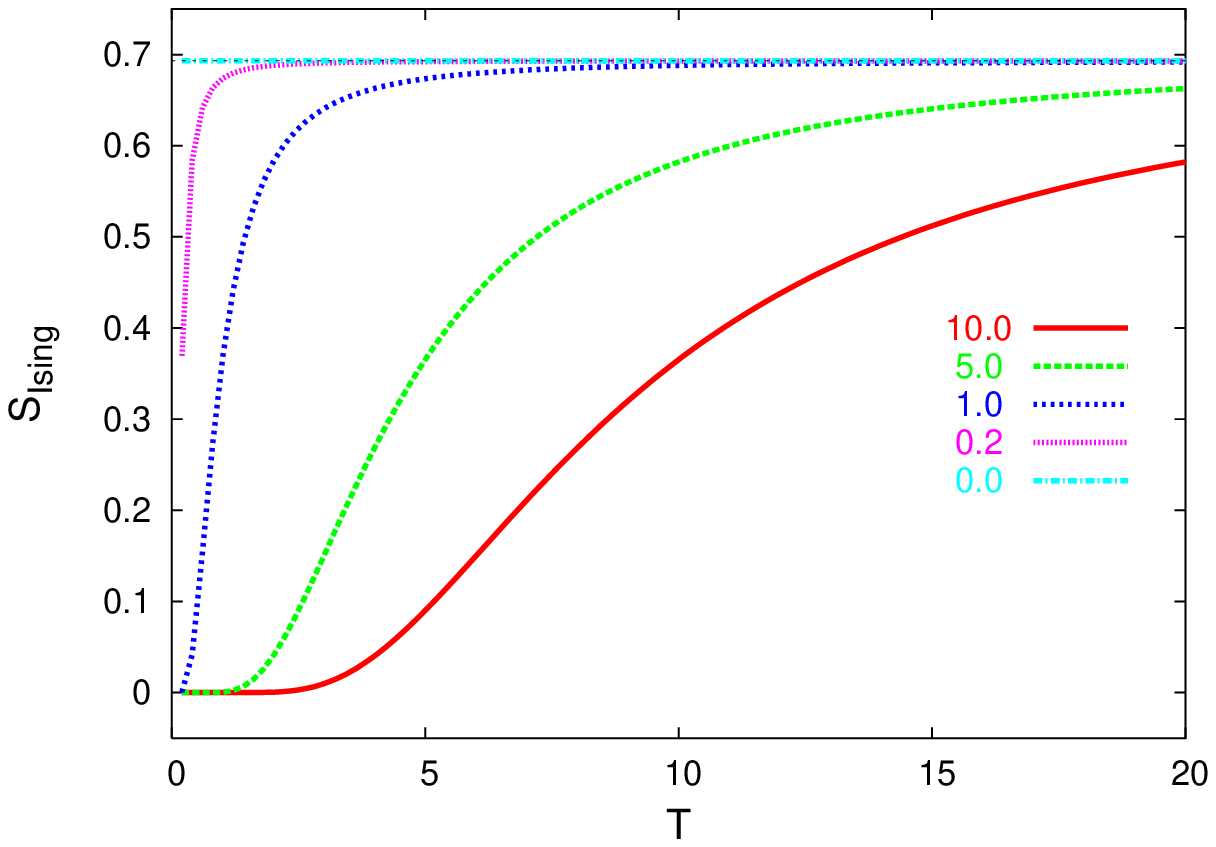}
\end{minipage}
\hskip -0.4cm
\begin{minipage}[t]{4.cm}
\includegraphics[width=8.cm]{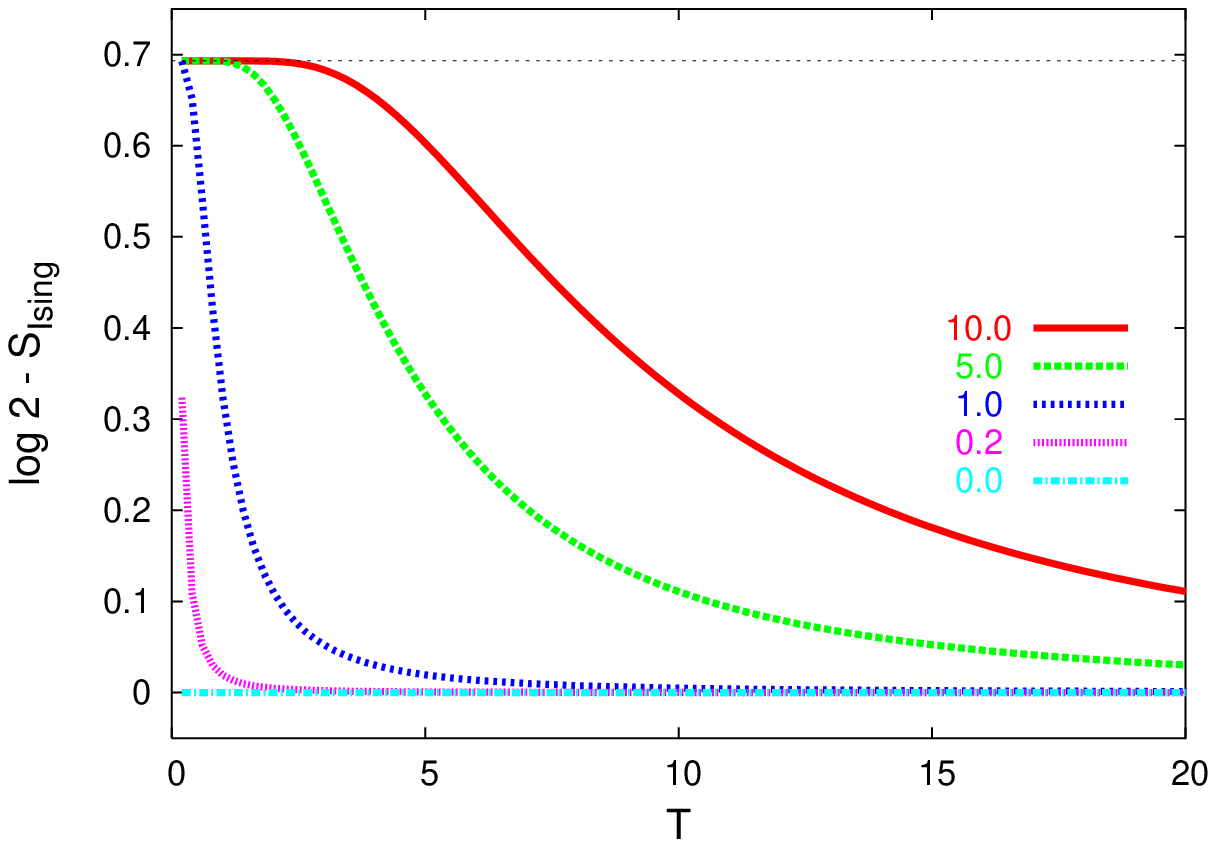}
\end{minipage}
\caption[Fig. 3]{\label{fig:3}\sf \footnotesize
    The entropy for 1D Ising model without external field calculated from
    the canonical Hamiltonian, Eq.~\ref{eq:Isingham}, depicted in
    dependence upon the temperature $T$ and for different values of $J$, the
    exchange coupling. Two asymptotic regions exist here. They are
    $S\rightarrow 0$ and $S\rightarrow \ln 2$ for temperatures
    $T\rightarrow 0$ and $T\rightarrow \infty$, respectively. In the
    right panel we depict the entropy difference from the $SU(2)$ ground
    state entropy, $\ln 2$, (see text). This entropy difference
    (Eq.~\ref{eq:IsingSdif}) represents the ground state entropy for the 1D
    Ising model at finite temperature.}  
\end{figure}

     The simplest special case of the electron-electron interaction is 
the Ising nearest neighbor chain interaction, for which only the spin
matrices in the z-direction \hbox{$s^z_i$ ($\equiv \sigma^z$ Pauli spin
  matrix)} and $s^z_{i+1}$ are present. Then the interaction Hamiltonian is simply

\begin{equation}
\mathcal{H}_{Ising} = -J \; \sum_{i}{s}_i^z {s}_{i+1}^z
                      -h \; \sum_{i}{s}_i^z
\label{eq:Isingham}
\end{equation}

\noindent
where $J$ is the simple nearest neighbor coupling and $h$ is the external
magnetic field. The interaction in Ising model 
is a completely classical since the spin operators
${s}^z_i$ can be replaced simply by their diagonal values. The
$SU(2)$ group structure ordinary reduces to its center $Z(2)$. Therefore,
in the ground state maximum two states are expected. In other words, each
spin can equally have one of the two directions up and down. 

The entropy per spin at finite temperature for $h=0$ reads
\begin{eqnarray}
\frac{S(T)}{N} &=& \ln \left[1+e^{-2J/T}\right] + \frac{J}{T}
                         \left[1-\tanh\left(\frac{J}{T}\right)\right] 
                   \label{isingST}
\end{eqnarray}
We see these results in Fig.~\ref{fig:3}a, where we have purposely taken
the energy scale for $J$ to compare with $S_{q,2}$ in Fig.~\ref{fig:1}a. It
is clear that the entropy is vanishing for vanishing temperatures but for
high temperature, $S\rightarrow \ln 2$. This
means that the ground state of Ising model does not reflect the structure
of $SU(2)$ symmetry group. In $SU(2)$ there are two states even at zero
temperature, which leads to one spin entropy equals to $\ln 2$. Therefore, the
canonical results in Fig.~\ref{fig:3}a indicates the fact that the ground state
structure in the Ising model is not included in the grand canonical
partition function. It is clear that the deviation from the characterized
$SU(2)$ structure is strongly depending upon the exchange coupling,
$J$. For all temperatures but only if $J=0$, the entropy per spin always
equals to $\ln 2$.

In doing a comparison with our models given in
section~\ref{subssu2}, we are left with an ad~hoc inclusion of an
additional part of the entropy reflecting the ground 
state structure in the Ising model. Therefore, we recall the structure of
the Heisenberg model, for which we can write down a wavefaction for the
strong correlated spin matrices $\sigma^z$ at $T=0$. Thus the density
matrix for the two possible states \hbox{$|0>,\; |1>$} which are
corresponding to the two spin directions \hbox{$|\uparrow>,\;
  |\downarrow>$}, respectively, reads     
\begin{eqnarray}
\rho = \sum |\Psi>\;P_{\Psi}\;<\Psi| &=&
\frac{1}{\sqrt{2}}\left(|0>|0^*>+|1>|1^*>\right)\nonumber \\
& & \frac{1}{\sqrt{2}}\left(<0|<0^*|+<1|<1^*|\right), \label{isingrho}
\end{eqnarray}
where $P_{\Psi}$ is the probability of each eigenstate $\Psi$. From the
traceable E.~\ref{isingrho} we get the reduced density matrix by projecting out the
conjugated components. 
\begin{eqnarray}
\rho^r= \Tr_b\; \rho &=& \frac{1}{2}\left(|0><0|+|1><1|\right) \label{isingrrho}
\end{eqnarray}
Obviously, the probability of each spin direction and correspondingly the
two eigenvalues from the reduced density matrix, Eq.~\ref{isingrrho}, are equal
($\lambda=1/2$). Plugging into Eq.~\ref{eq:enteig}, results in the entropy
for single spin of 1D Ising model at zero temperature, $S=\ln
2$. In getting this value, we obviously assumed that $J=0$ and considered only the
spin {\it eigen}fluctuation in the ground state. i.e, quantum states.

As we have done in section~\ref{subssu2} (Eq.~\ref{eq:denmattem}), we
subtract from this value the $T$-depending entropy part resulting in the
temperature dependence of the ground state (quantum) entropy 
\begin{eqnarray}
S(T) &\equiv& \ln 2 - S_{Ising}(T) \label{eq:IsingSdif}
\end{eqnarray}
These results are given in Fig.~\ref{fig:3}b. At $T=0$,
the entropy starts from the value $\ln 2$. With increasing temperature the
ground state entropy of the 1D Ising model monotonically decays. It reaches
its asymptotic zero value for high 
$T$. Qualitatively, we got the same results for our 
models given in section~\ref{subssu2} and graphically illustrated
in Fig.~\ref{fig:1}~and~\ref{fig:2}. Obviously, we left arbitrary the units
of the energy 
density in the Boltzmann term in Eg.~\ref{eq:denmattem} and of the exchange
coupling in Eq.~\ref{isingST}.

\subsection{Quantum $XY$ spin chain}

\begin{figure}[thb]
\vskip -.2cm
\begin{minipage}[t]{80mm}
\includegraphics[width=8.cm]{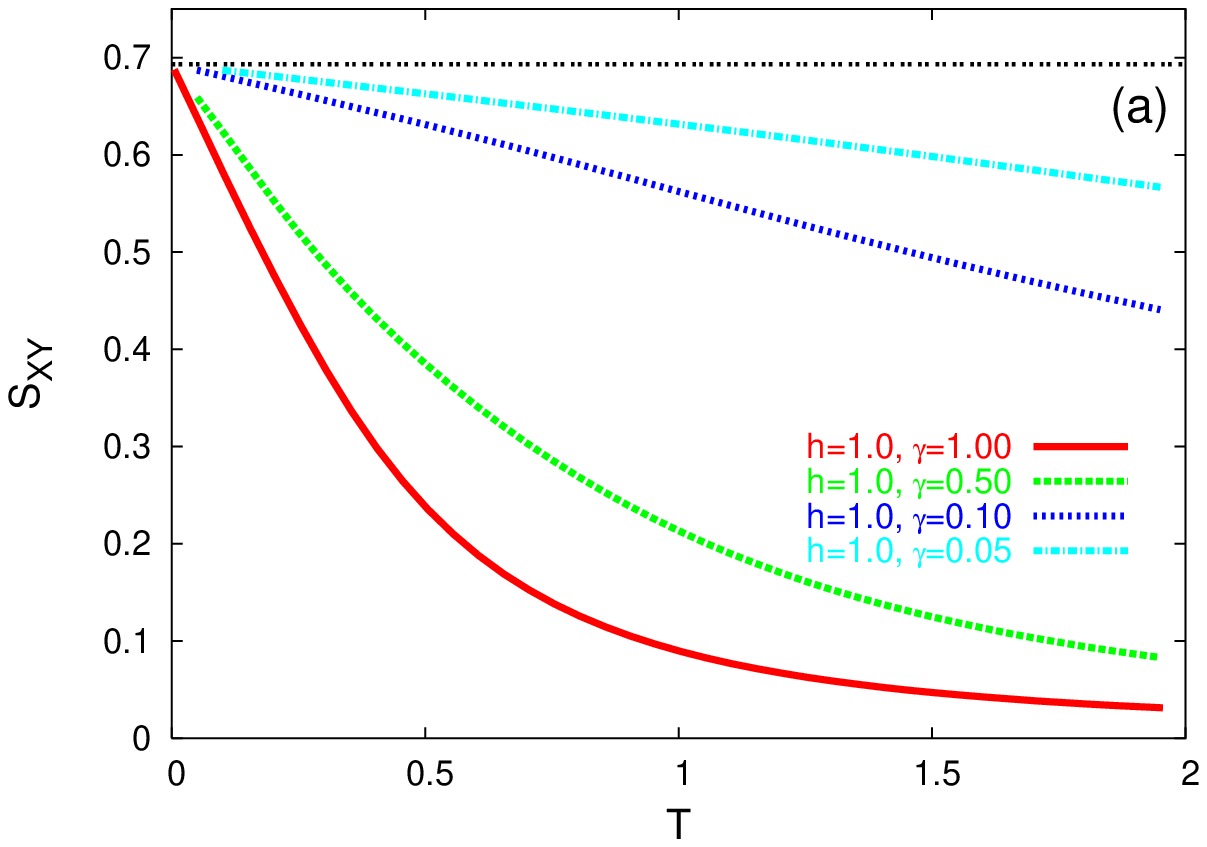}
\end{minipage}
\begin{minipage}[t]{4.cm}
\includegraphics[width=8.cm]{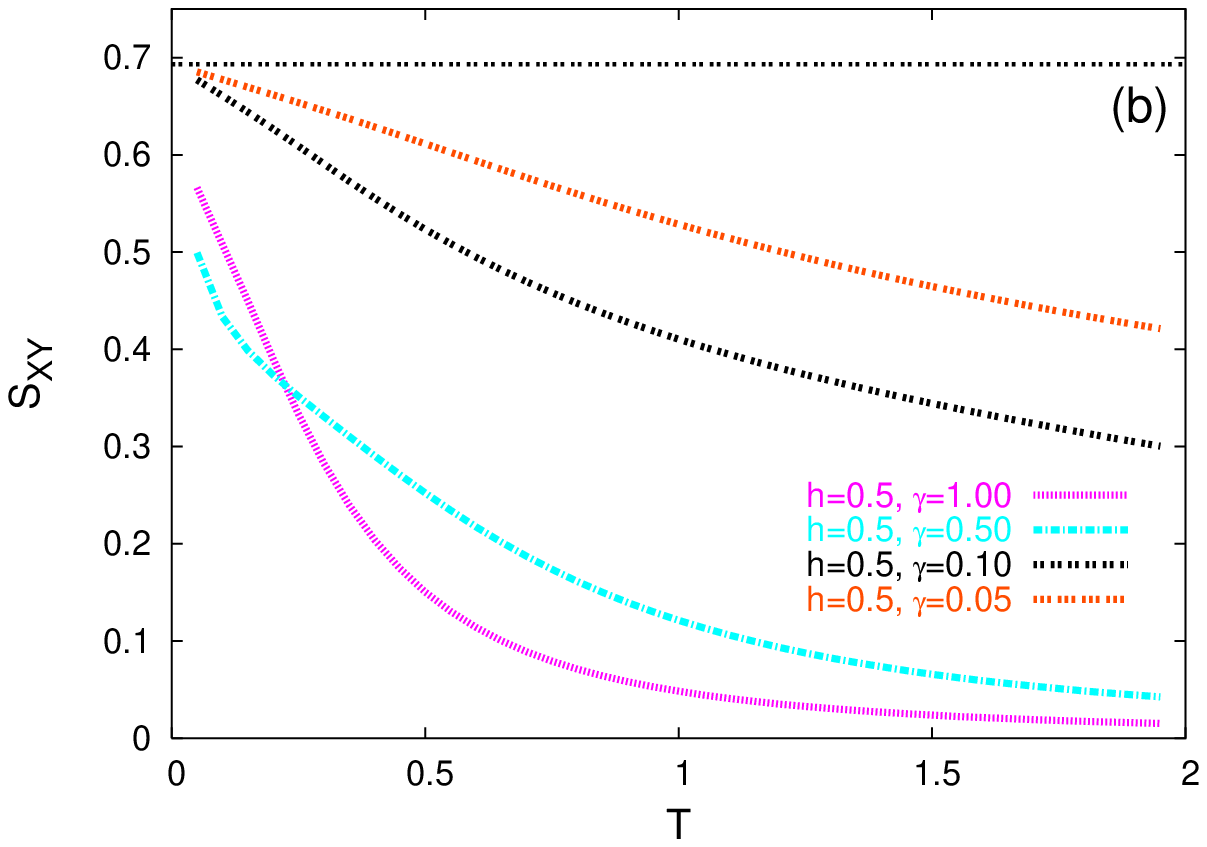}\\
\end{minipage}
\vskip -0.5cm 
\begin{minipage}[t]{4.cm}
\hspace*{4.cm}
\includegraphics[width=8.cm]{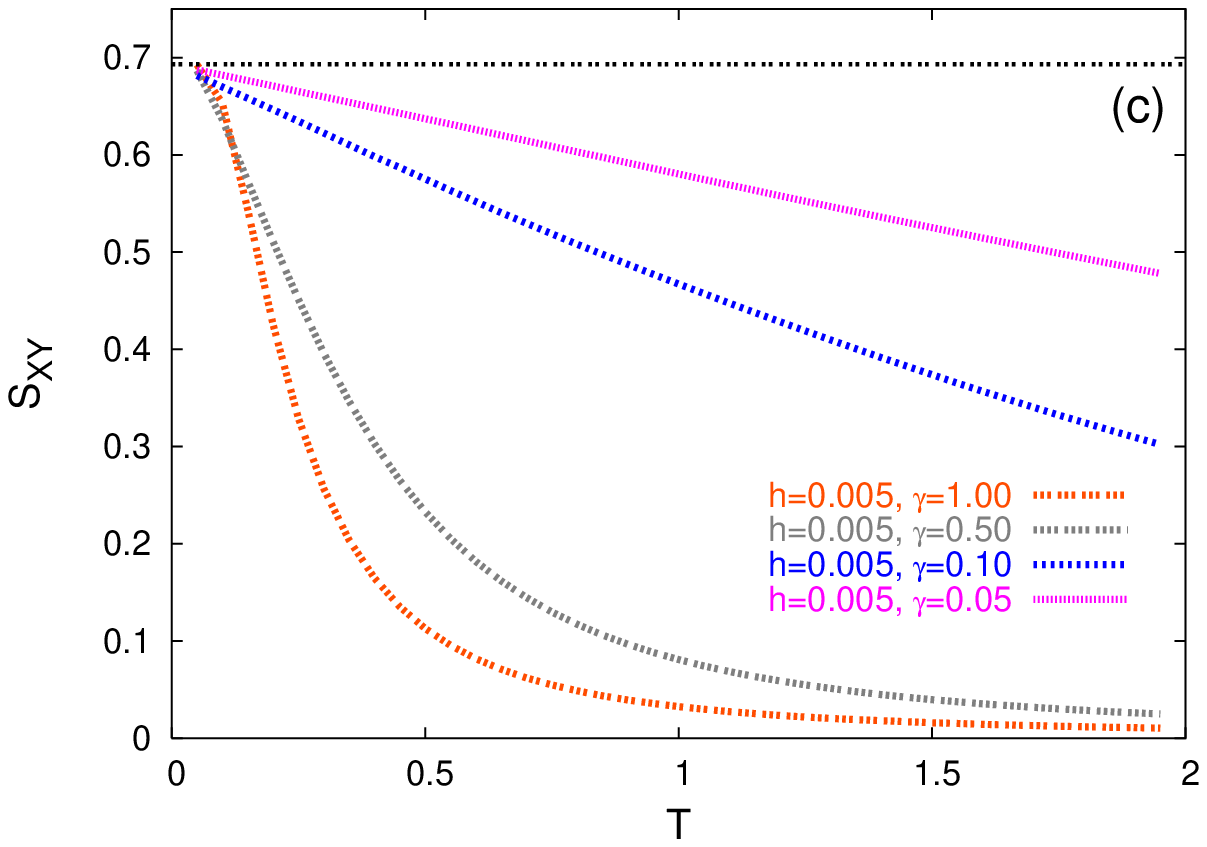}
\end{minipage}
\vskip -0.1cm 
\caption[Fig. 4]{\label{fig:4}\sf \footnotesize
  Entropy per spin site calculated for different temperatures for
  $XY$-model. (a) gives the results for constant $h=1.0$ but different
  $\gamma$ values (see text). We note that with increasing $T$ the entropy
  $\mathcal{S}$ decreases exponentially and the $T$-range within which
  $\mathcal{S}\rightarrow 0$  
  becomes wider with smaller $\gamma$ values. Also deceasing $\gamma$ increases
  the finite entropy and the system favors going toward the ground state value,
  $\ln 2$. Almost the same behavior can be seen in the plots (b) for $h=0.5$
  and (c) for $h=0.005$. 
}
\end{figure}

    Another exactly solvable model relating to the Heisenberg model is the
\hbox{$XY$ model~\cite{LSM,Kat}}. In one spatial dimension it is a highly  
correlated quantum mechanical system also in the ground state~\cite{BMD}.
Furthermore, it is known~\cite{vidal,BMD,Henkel} that the $XY$-model in the
ground state has a very complex structure containing an oscillating region for small
values of $h$ and $\gamma$ in addition to the expected ferromagnetic
and paramagnetic phases for larger values of these parameters. Also,
for a class of time dependent magnetic fields the magnetization has been
shown to be nonergodic~\cite{BMD}. Recently, the correlations of the 
spins  has been calculated using the reduced density matrix~\cite{vidal}
in order to find the ground state entropy for a block of spins yielding
the expected logarithmic behavior. 

\noindent
The Hamiltonian can be written in terms of the spin components in the
$x$ and $y$ spin-directions with an external field in the $z$-direction
\begin{equation}
{\mathcal{H}_{XY}} = \frac{1}{2}\sum_{i}\left[ (1+\gamma) s^x_i s^x_{i+1} + 
                     (1-\gamma) s^y_i s^y_{i+1}\right] - h \sum_{i} s^z_i
\label{eq:XYham}
\end{equation}
\noindent
where $\gamma$ is the spin anisotropy parameter in the $x$ and $y$ spin-directions.
The exact solution of this model has been carried out long ago~\cite{LSM,Kat}.
The spins are allowed to take an arbitrary angle,  $k\in[0,2\pi]$ with
cyclic boundary conditions. By using the fermionic 
variables, $a_k^{\dagger}, a_k$ the Hamiltonian\footnote[3]{The positive
  sign of Hamiltonian refers to attractive nearest neighbor spin
  interactions. By rotating the chain along the spin $z$ direction to every
  second spin one can flip this sign \cite{vidal}.}  may be
written as
\begin{eqnarray}
H &=& \sum_k \Lambda_k \left(a_k^{\dagger} a_k -\frac{1}{2}\right) \\
\Lambda_k^2 &=&
\frac{1}{\gamma^2}\left[(h-1)^2+4\left(h-1+\gamma^2\right)sin^2\left(\frac{k}
{2}\right) - 4\left(\gamma^2-1\right)sin^4\left(\frac{k}{2}\right)\right]
\end{eqnarray}
\noindent
Then the entropy per spin site at finite temperature reads
\begin{eqnarray}
\frac{S_{\gamma}(T)}{N} &=& \frac{1}{2\pi} \left[ \int_{0}^{\pi/2} dk \;
                            \ln(1+e^{-(\Lambda_k-\mu)/T}) +
                            \frac{1}{T}\int_{0}^{\pi/2} dk \; 
                            \frac{e^{\Lambda_k/T}(\mu-\Lambda_k)}{(e^{\Lambda_k/T}
                             + e^{\mu/T})} \right]
\label{eq:XYEntop1}
\end{eqnarray}

\noindent
The entropy per spin given in Eq.~\ref{eq:XYEntop1} represents 
the change of the mixing of the finite temperature state form the
ground state due to the thermal effects. Since the quantum Ising model with
$\gamma=1$ has always the smallest value of $\mathcal{S}_{XY}$,  
the effects of spin mixing disappear at the lower temperature.
From these plots we see that ${\mathcal{S}_{XY}}$ yields results
analogous to our models for $SU(2)$ and $SU(3)$ discussed in section 3,
depending upon the units of energy density, coupling, and the parameters,
$h$ and $\gamma$. 
In Fig.~\ref{fig:4} we plot the entropy per spin site versus the
temperature $T$ and for different value of $h$ and $\gamma$. Depending on
these parameters the results of $XY$ model given in
Fig.~\ref{fig:4} can be compared with that of the Ising model,
Eq.~\ref{fig:3}. The condition is that $\gamma=1$ and $h\rightarrow 0$. We
should notice the arbitrary units in both figures. For
$\gamma\rightarrow 0$ the value of the ground state entropy favours to stay
constant for all temperarutes. The same results are found in 1D Ising model
for the limits $J\rightarrow 0$. We
can conclude that the ground state entropy for the $XY$ spin chain are
qualitatively comparable with our models for the $SU(2)$ and $SU(3)$.

\section{Discussion}

We have calculated the entropy for colored quark states in the hadron singlet 
and octet structures and extended our considerations to finite temperatures. 
According to the third law of thermodynamics the pure states, 
such as completely specified enclosed hadronic systems,
possess a vanishing entropy at very low temperatures. 
Therefore, the hadron bags as macroscopic isolated objects are
expected to have zero entropy at a vanishing temperature. 
However, if we further consider the {\it microscopic} hadron {\it
  subsystems} at low temperatures up to the quark mass, 
the quantum entropy is seen to have a finite value. It is a curious fact
that this value is usually simply ignored or approximately taken 
to be zero especially for classical systems, 
although it had been already recognized since the nineteenth
century~\cite{Planck64}  
and confirmed for the laws of quantization of statistical states for the
thermodynamics in last century~\cite{vNeu,LaLi,Schr}. 
Therefore, we think that there is no compelling reason to
assume that the value of the ground state entropy of quantum or even 
classical {\it subsystems} is zero, just because one 
believes that the system to which they are assigned is
enclosed and thereby must have a vanishing entropy. The {\it subsystems} have
other degrees of mixing and thereby a finite entropy, even if the whole
closed system is isolated and consequently has a vanishing entropy.
Nevertheless, the ground state entropy is important and therefore should be
taken into consider at low temperatures. As we have seen the upper limit of
temperature of the validity or the importance of the ground state entropy
is characterized by the quark mass, i.e., $T\in[0,m]$. For light quarks and
if we are 
interested on the QCD phase transition or on the quark matter at very high
temperatures, the ground state entropy of colored quarks is no longer
significant. On the other hand, in the interior of stellar compact objects,
cold dense quark matter is highly expected. for which the ground state entropy
would play an important role, especially, on understanding the
superconductivity on cold dense quark matter~\cite{Rischke} and the phase
transition from neutron to quark matter in the hybrid stars and the stability and
structure of these compact stars~\cite{David,Igor}. \\
~\\
\indent
The quark and antiquark 
mathematically build up the Schmidt decompositions of meson-state, where the
Schmidt numbers simply represent the normalization of their
wavefunctions. Furthermore, from the quantum teleportation we know that the 
colored quark and antiquark can be considered as mutual 
purifications for each other. Each single state is equally weighted in
the decomposition of meson states, from which each state possesses an equal
probability. On the other hand each single quark state represents a certain
degree of mixing and therefore has a finite entropy although the meson state
 -- as a pure isolated state -- must have zero entropy. The baryon states have
the doubly reduced density matrix for each single quark state appearing
twice. The reduced density matrix give the spatial mixing of
{\it subsystems}. The resulting quantum entropy at zero temperature gives 
the maximal quantum entropy for completely mixed states.

     As we have seen the octet states are much more complex 
than the singlet states, since the octets have many more and quite
different states. 
When we looked at the Pauli matrices we realized that the eighth
Gell-Mann matrix, $\lambda_8$, counts also the other states and generally
no one of them 
can be considered as a purification of the other states. At finite temperature the
octet structure becomes much more complex. The reason for this is that with
increasing temperature the correlations between the states become stronger,
and at high temperature the correlated states reach the asymptotic
value. At high temperatures the states of the Pauli type become more
probable. Furthermore, the pure ground state becomes no longer possible. 
Thus only the unoccupied states are the most available at high temperature. \\ 
~\\
     We have postulated simple models for the thermal dependence of the ground
state entropy. We saw that the singlet state is a completely mixed state
with the maximum value of the entropy given by $\ln 3$ at vanishing
temperature. We have used a Boltzmann-like factor for the thermal dependence
of the entropy. With increasing temperature the reduction of the ground state
entropy is still continuing. At high temperature the mixing and consequently
the entropy vanish entirely. The octet states show qualitatively the same
results. The behavior of octet state with the temperature reflects the complexity 
of their basic structure. We have noted that by decreasing the quark masses the
range of temperatures becomes limited for reaching this asymptotic region.  
Since the asymptotic value at high  temperatures has remained more
than three quarters  of its ground state value of $\ln 3$, there are still
considerable correlations between two of the three color states. This
observation is based on the complex structure of $SU(3)$. Obviously, the root
structure forbids a complete cancellation of the real solutions which is
needed to maintain the trace condition on the reduced density matrix. Thus
two states are always matched against one, which forms the 
subsystem. Hence in the high temperature limit one color 
vanishes while the other two remain mixed and thereby correlated. We
can conclude that ground state favors the color singlet state with complete 
mixing, meanwhile $SU(3)$ -- in the high temperature limit -- favors the
octet states involving mostly two colors. This situation for $SU(3)$ can be
contrasted with $SU(2)$ where the triplet and the singlet states have the
same reduced density matrices except for the pure triplet states. \\
~\\ 
\indent 
     In order to further model the ground state structure and the entropy
     of $SU(2)$,  
we utilized some known {\it classical} and {\it  quantum} spin chains,
which have strong correlations at finite temperature. We investigate the
behavior of their Hamiltonians at finite temperature. The simplest and
widely used {\it classical} spin model is the one-dimensional Ising model.  
The other solvable model which is related to Heisenberg model is 
$XY$ spin chain. It is a highly correlated quantum mechanical system in the
ground state. One of the most valuable results we have gotten 
from our investigation here is that the classical spin model is not able to
successfully  
describe the ground state. The entropy from the partition function is simply
zero at zero temperature. Thus we plot the entropy differences of the
finite temperature states to the ground state. If we additionally include 
thermal terms similar to those in our models for $SU(2)$ and $SU(3)$ the entropy
starts from zero and monotonically increases up to the asymptotic value of
$\ln2$. Also we have found that this asymptotic limit does not depend
on the spatial dimension. Particularity with regard to the $XY$-model it
has a nontrivial structure depending on the parameters of spin asymmetry and
external field, which has been recently investigated through the correlation
functions in the ground state~\cite{vidal}.

\medskip
\section{Conclusions and Outlook}

In this work we have compared the quantum definitions as contrasted to the 
classical concepts of entropy in relation to the temperature. We have noticed  
that in general the quantum definition is important in the 
low temperature limit, while the classical concepts usually relate
to higher temperatures. First we have discussed the ground state
entropy, which is strictly a quantum definition and does not itself appear in
classical physics. We have evaluated and contrasted the symmetry structure
for both the $SU(2)$ and $SU(3)$ color groups. For these symmetries we have 
used simple thermodynamical models involving the color ground state entropy
to show how the quantum mixing entropy disappears for the entropy 
differences with increasing temperature. From this result we have also 
discussed how this disappearance can give rise to new pure states 
in the high temperature limit. \\
~\\
\indent
     One motivation for this work has been to study the correlation
between the quarks and antiquarks. In recent QCD lattice simulations
one can find indications of ground state behavior at short distances
and low temperatures~\cite{Felix1,Felix2,FelixPhD}. These results could very well 
relate to the entropy we described above. 
As a next step we want to look into the thermodynamical properties 
of these correlations in relation to the present study~\cite{MiTa1}.
There are many application of the finite entropy of colored quarks at zero
and very low temperatures much below the temperature of QCD phase
transition from hadron to quark-gluon plasma~\cite{Taw1,Taw2}. 
This endeavor could help demonstrate the usefulness of the quantum
entropy in the description of the thermal properties of the strong 
interaction at very low temperatures. Furthermore, in two outcoming works
we shall include the effects of the gluons and the chiral symmetry in the
future. We will also consider the effects of finite value of 
entropy at low temperature on the pressure inside the hadron
bag~\cite{MiTa2}. Also the effects of the quantum entropy on the
condensates of quark pairs with strong correlations and at very low
temperatures and very high quark chemical potentials~\cite{MiTa3}.\\ 
~\\
\indent
     Thus we have seen in several different models 
how the usual thermodynamical entropy gotten 
from the evaluation of the partition function 
acts as a means of disorganizing the ground state.
It undoes the entanglement in some cases completely and
in others only partially. Therefore it has the effect
of lowering the correlations between the quarks and 
antiquarks seen in the ground state. \\

\medskip
\noindent{\bf\Large Acknowledgments}

\medskip
The authors would like to thank Frithjof~Karsch, Krzysztof~Redlich and
\hbox{Helmut~Satz} for the very helpful discussions. D.E.M. is very grateful to the
Pennsylvania State University Hazleton for the sabbatical leave of absence
and to the Fakult\"at f\"ur Physik der Universit\"at Bielefeld.

\end{document}